\documentclass[seceq,preprint]{ptptex}

\usepackage{graphicx}

\newcommand{\bm}[1]{\mbox{\boldmath${#1}$}}
\renewcommand{\bm}[1]{\boldsymbol{#1}}

\newcommand{\Eq}[1]{Eq.~(\ref{#1})}
\newcommand{\eq}[1]{(\ref{#1})}

\newcommand{\VEV}[1]{\left\langle{#1}\right\rangle}
\newcommand{\order}[1]{{\cal O}\left({#1}\right)}

\newcommand{\fun}[1]{\!\left(#1\right)}
\newcommand{\abs}[1]{\left\vert{#1}\right\vert}

\def\wb#1{\vbox{\ialign{##\crcr%
          \hskip 1.0pt\hrulefill\hskip 0.3pt%
          \crcr\noalign{\kern-1pt\vskip0.07cm\nointerlineskip}%
          $\hfil\displaystyle{#1}\hfil$\crcr}}}

\def\leftB{\Bigl}
\def\rightB{\Bigr}

\newcommand{\ie}{\textit{i}.\textit{e}.}
\newcommand{\disp}{\displaystyle}

\newcommand{\lsim}{ \mathop{}_{\textstyle \sim}^{\textstyle <} }

\renewcommand{\L}{{\rm L}}
\newcommand{\R}{{\rm R}}

\newcommand{\GeV}{{\rm GeV}}
\newcommand{\TeV}{{\rm TeV}}

\newcommand{\MPl}{M_{\rm Pl}}
\newcommand{\Mc}{M_{\rm c}}
\newcommand{\Nc}{N_{\rm c}}
\newcommand{\Nf}{N_{\rm f}}

\newcommand{\GSC}{G_{{\rm SC}}}
\newcommand{\GSM}{G_{{\rm SM}}}
\newcommand{\Ncb}{\wb{N}_{\!{\rm c}}}

\preprintnumber[3.5cm]{
KANAZAWA-04-18\\[-1mm]
KUNS-1943\\[-1mm]
NIIG-DP-04-2\\[-1mm]
November, 2004}

\title{
Large Mass Scale by Strong Gauge Dynamics\\
with Infrared Fixed Point
}

\author{
Tatsuo~{\sc Kobayashi}\rlap,${}^{1,}$\footnote{
	E-mail: kobayash@gauge.scphys.kyoto-u.ac.jp}
Hiroaki~{\sc Nakano}\rlap,${}^{2,}$\footnote{
	E-mail: nakano@muse.sc.niigata-u.ac.jp}
Haruhiko~{\sc Terao}\rlap\,${}^{3,}$\footnote{
    E-mail: terao@hep.s.kanazawa-u.ac.jp}\\
and
Yoshihisa~{\sc Yamada}${}^{4,}$\footnote{
	E-mail: yoshi@muse.sc.niigata-u.ac.jp}
}

\inst{
  $^{1}$\textit{%
Department of Physics, Kyoto University, Kyoto 606-8502, Japan}\\
  $^{2,4}$\textit{%
Department of Physics, Niigata University, Niigata 950-2181, Japan}\\
  $^{3}$\textit{%
Institute for Theoretical Physics, Kanazawa University, 
Kanazawa 920-1192, Japan}
}

\abst{
We consider a mechanism for realizing the desired decoupling
of strongly-coupled sector which is supposed to generate 
hierarchical structure of the Yukawa couplings.
In our mechanism, the same strongly-coupled sector
is responsible for generating a sufficiently flat potential and 
a large vacuum expectation value (VEV) of a gauge-singlet scalar field
by suppressing its soft scalar mass and self-coupling.
Vacuum instability is caused by supersymmetry-breaking 
A-term of order 10 TeV.
We explicitly demonstrate the infrared convergence of 
soft scalar masses due to strongly-coupled dynamics
and show the soft mass of the singlet 
is at most comparable to soft masses of squarks and sleptons,
which are much suppressed than the A-term.
The physical mass scale of the decoupling
is calculated in a self-consistent way.
We also reinterpret the result
in terms of a RG-improved effective potential.
}

\begin{document}

\maketitle

\section{Introduction}
\label{sec:intro}

An attractive candidate for models beyond the Standard Model (SM) 
is provided by softly-broken supersymmetry (SUSY). 
In general, soft SUSY-breaking parameters can be 
a new source of flavor mixings and CP violation, and 
search for such effects may give us an indication of new physics.
However, an arbitrary set of SUSY-breaking parameters 
results in too large flavor violations.
One way to suppress SUSY flavor violating effects is to assume that 
sfermions have diagonal and degenerate masses at some high-energy scale.
{}For instance,
the flavor-blind mediation mechanism of SUSY breaking 
has been discussed extensively 
in the context of extra dimensions. 
In this approach,
SUSY flavor violations are suppressed
irrespectively of the origin of hierarchical structure of 
Yukawa couplings and mixing angles.

An alternative approach to SUSY flavor problem 
was proposed by Nelson and Strassler \cite{NS1}
and has been developed in Refs.~\citen{KT:2001,NS2,Terao,KNNT1,transfer},
by utilizing interesting properties of strongly-coupled gauge theories  
with nontrivial infrared (IR) fixed point.\cite{BZ,Seiberg} \ 
When such a superconformal field theory (SCFT) is perturbed 
by soft SUSY-breaking terms, these soft terms become suppressed 
in IR regime \cite{SQCD_convergence}.
Specifically soft scalar masses satisfy IR sum rules \cite{KT:2001}:
\begin{eqnarray}
 \sum_{i} T_{i}^{a}m^2_i
 \longrightarrow 
 0 \ , \qquad
 m^2_i + m^2_j + m^2_k
 \longrightarrow 
 0 \ ,
\label{sumrule}
\end{eqnarray}
where $m_i$ is a soft scalar mass of matter chiral field $\Phi_i$
and $T_i^a$ is Dynkin index of $\Phi_i$ under the gauge group $G_a$.
[These IR sum rules correspond to the fixed point condition of
the gauge coupling and a superpotential coupling 
$\lambda_{ijk}\Phi_{i}\Phi_{j}\Phi_{k}$, respectively.]
It was then argued \cite{NS1,KT:2001} that
one can construct a model in which these IR sum rules 
are powerful enough to constrain squarks/slepton masses.
Consequently one can obtain an approximately degenerate
sfermion masses at weak scale,
independently of mediation mechanism of SUSY breaking.

The hierarchy of SM Yukawa couplings can also be generated 
by coupling SCFT's to SUSY SM (or its extension).
Through an interaction with SCFT sector,
SM matter fields gain large anomalous dimensions $\gamma_i$ 
so that the SM Yukawa couplings $y_{ij}$ obey power-law running,
\begin{eqnarray} 
 y_{ij}\left(\mu\right)
 &\approx&
 y_{ij}\left(\Lambda\right)
  \left(\frac{\mu}{\Lambda}\right)^{%
  \frac{1}{2}\left(\gamma_i+\gamma_j\right)
  } \ ,
\label{powerlaw:yukawa}
\end{eqnarray}
where $\Lambda$ is a scale below which strongly-coupled sector 
may be regarded as approximately conformal.
In the original Nelson-Strassler scenario, 
family-dependent anomalous dimensions give rise to 
the hierarchy of SM Yukawa couplings.
[Alternatively, Yukawa hierarchy transfer scenario
was proposed \cite{transfer}, where 
one can achieve complete degeneracy of sfermion masses
by combining the SCFT idea either with $U(1)$ flavor symmetry 
or extra-dimensional setup.]

Now, it is important to notice that
power-law running \eq{powerlaw:yukawa} of Yukawa couplings
should be terminated at a certain scale $\Mc$.
In other words, SCFT sector should decouple from SUSY SM-sector.
Such decoupling is possible 
if `SCFT' sector is not an exact conformal field theory;
Some matter fields have a tiny bare mass term 
at the ultraviolet (UV) cutoff scale.
If the mass term is tiny enough,
the perturbed theory can be regarded as approximately conformal.
However, as we go down to lower energy scale,
the mass term becomes important and eventually
the `SCFT' sector becomes massive.
On phenomenological ground \cite{NS1,KT:2001,NS2},
the decoupling scale $\Mc$ is required to be 
of several order below the scale $\Lambda$.
If we identify $\Lambda$ with the Planck scale $\MPl$,
the required decoupling scale is around
an intermediate scale $10^{13}{\rm -}10^{15}\,\GeV$.

Our aim here is to discuss an origin of the required mass term.
One may add an invariant mass term (by hand)
if smallness of that mass parameter
is related to smallness of other ones.
{}For instance, Ref.~\citen{KT:2004} discussed 
an application of SCFT to SUSY Higgs sector,
where its decoupling is realized by an invariant mass term
whose origin is related to that of the $\mu$ term in SUSY SM.
On the other hand, 
there are several ways for generating it dynamically.
One way is to assume another strong gauge dynamics
that is responsible for mass generation,
as was already mentioned in Ref.~\citen{NS1}.
Another possibility is to use an anomalous $U(1)$ gauge symmetry
which generates a nonvanishing Fayet-Iliopoulos term.
In such a model, a charged scalar field $\chi$ often develops
the VEV $\VEV{\chi}$ that cancels the Fayet-Iliopoulos term.
If matter fields in SCFT sector are charged under the $U(1)$
and have a proper coupling to the $\chi$ field,
then they obtain mass terms that are suppressed
by a power of $\VEV{\chi}$.
This possibility was discussed in Ref.~\citen{Dproblem}
in the context of gauged $U(1)$ flavor symmetry.

In the present paper, we propose a simpler possibility; 
SCFT sector becomes massive and decouples from SM-sector
via its own superconformal dynamics.
We also give a rough estimate of how large mass scale
can be generated by our mechanism.
In the next section, 
we set up a toy model which we use to explain our idea.
Specifically we introduce a massless singlet field $S$
that couples to SCFT sector through a renormalizable superpotential.
In \S\ref{sec:RGE}, we study 
renormalization group (RG) evolution of parameters in the model.
The cubic self-coupling of $S$ is suppressed
due to a large anomalous dimension generated by SCFT dynamics,
while the corresponding $A$ parameter is \textit{not}.
We also demonstrate explicitly that
soft scalar mass of the singlet $S$ becomes suppressed
and converges on a small value.
In \S\ref{sec:mechanism},
we show how an approximately flat potential for $S$ is generated
by strongly-coupled dynamics of SCFT.
Consequently we can generate a hierarchically large VEV compared
with the size of soft SUSY-breaking parameters.
This is our main result.
In addition, we also discuss RG-improvement of the effective potential.
It turns out that
we obtain apparently different expressions for the scalar VEV
from the tree potential with RG running parameters substituted
and from the RG-improved potential.
We explain that such discrepancy disappears
if we calculate the `physical' decoupling scale $\Mc$
defined in a self-consistent manner.
Appendix is devoted to an illustration of IR convergence of soft scalar 
masses by using one-loop anomalous dimensions.

\section{Model}
\label{sec:model}

In this section, we describe a model
with an IR fixed point.
{}For simplicity, we discuss a toy model with left-right symmetry.

We consider a $\GSC=SU(\Nc)$ gauge theory with $\Nf=2n$ flavors,
which we will refer to as SC sector. 
In addition, we add several singlets under $\GSC$.
The matter content is shown in Table~\ref{tab:matter}.
The model has $\GSM\equiv{}SU(n)_\L\times{}SU(n)_\R$ flavor symmetry
which we weakly gauge.
In a realistic application,
$\GSM$ will be regarded as the SM gauge group 
(or its 
extension)
and $\psi_{\L,\R}$ as quark and lepton fields.
So we will refer to $\GSM$ as SM gauge group 
and to $\psi$'s as SM-matter fields also in this toy model.
In addition
we impose $Z_2$ symmetry which exchanges $\L$ fields with $\R$ fields.

\begin{table}[tb]
\caption{
Matter content ($\Nf=2n$)
}
\label{tab:matter}
\begin{center}
\begin{tabular}{r|cc|cc|cc|c}
 \hline \hline
${}^{\phantom{A^{A}}}\!\!\!\!\!\!\!$
 & $\Phi_\L$ & $\Phi_\R$
 & $\wb{\Phi}_\L$ & $\wb{\Phi}_\R$
 & $\psi_\L$ & $\psi_\R$ & $S$
\\\hline
${}^{\phantom{A^{A}}}\!\!\!\!\!\!\!$
$SU(\Nc)$  & $\bm{\Nc}$    & $\bm{\Nc}$    & $\bm{\Ncb}$   & $\bm{\Ncb}$
           & $\bm{1}$	   & $\bm{1}$      & $\bm{1}$
\\\hline
${}^{\phantom{A^{A}}}\!\!\!\!\!\!\!$
$SU(n)_\L$ & $\bm{n}$      & $\bm{1}$      & $\bm{1}$      & $\bm{\wb{n}}$
           & $\bm{\wb{n}}$ & $\bm{n}$      & $\bm{1}$
\\
$SU(n)_\R$ & $\bm{1}$      & $\bm{n}$      & $\bm{\wb{n}}$ & $\bm{1}$ 
	   & $\bm{n}$      & $\bm{\wb{n}}$ & $\bm{1}$
\\
 \hline
\end{tabular}
\end{center}
\end{table}

The superpotential we consider is given by 
$W=W_{{\rm mess}}+W_{{\rm dec}}$ with
\begin{eqnarray}
W_{{\rm mess}}\!\!
 &=& \lambda_\psi\leftB[
     \psi_\L\fun{\bm{\wb{n}},\bm{n}}
     \Phi_\L\fun{\bm{n},\bm{1}}
     \wb{\Phi}_\L\fun{\bm{1},\bm{\wb{n}}}
    +\psi_\R\fun{\bm{n},\bm{\wb{n}}}
     \Phi_\R\fun{\bm{1},\bm{n}}
     \wb{\Phi}_\R\fun{\bm{\wb{n}},\bm{1}}
     \rightB] \ ,
\label{W:mess}\\
W_{{\rm dec}}
 &=& \lambda_{S\,}S\fun{\bm{1},\bm{1}}\leftB[
     \Phi_\L\fun{\bm{n},\bm{1}}
     \wb{\Phi}_\R\fun{\bm{\wb{n}},\bm{1}}
    +\Phi_\R\fun{\bm{1},\bm{n}}
     \wb{\Phi}_\L\fun{\bm{1},\bm{\wb{n}}}
     \rightB]
    +\frac{\kappa}{3!}\,S^3\fun{\bm{1},\bm{1}} \ , \quad
\label{W:dec}
\end{eqnarray}
where we have indicated $SU(n)_\L\times{}SU(n)_\R$ quantum numbers.
SM-matter fields $\psi$ couple to SC-sector through 
the superpotential term \eq{W:mess}, 
which we call messenger interaction.
On the other hand, the first term in \Eq{W:dec} 
represents a coupling to gauge-singlet field $S$,
and becomes mass term of SC-sector matter fields 
$\Phi$ and $\wb{\Phi}$ once $S$ develops a nonzero VEV.
It is the main subject of the present paper to describe 
how a large VEV of $S$ can be generated
thanks to strong $\GSC$ dynamics itself.

With sufficiently many matter fields,
the $SU(\Nc)$ gauge theory resides in the so-called conformal window
and its gauge coupling $g_{{\rm SC}}$ reaches an IR fixed point,
supersymmetric version of Banks-Zaks fixed point \cite{BZ,Seiberg}.
At this point the SC-sector matter fields $\Phi$ and $\wb{\Phi}$ 
obtain a large negative anomalous dimension
$(\gamma_{\Phi})_*=(\gamma_{\bar\Phi})_*=-\left(3\Nc-\Nf\right)/\Nf$,
and so the messenger interactions \eq{W:mess} become relevant.
Then SM-sector matter fields $\psi$ develop a large positive 
anomalous dimension $(\gamma_{\psi})_*$,
and eventually the coupling $\lambda_\psi$ reaches a new fixed point.
This is the basic setup of models in Refs.~\citen{NS1,transfer}.
In addition, we assume that the same is true for 
the singlet coupling $\lambda_S$ in \Eq{W:dec}:
\begin{eqnarray} \label{eqn:beta_lambdaS}
 \mu\frac{d}{d\mu}\ln\lambda_S
 =
 \frac{1}{2}\left(
 \gamma_S+\gamma_{\Phi}+\gamma_{\bar\Phi}
 \right)
 \longrightarrow{} 0 \ . 
\end{eqnarray}
Thus the singlet field $S$ also gains a large anomalous dimension
$(\gamma_S)_*$.

As was shown in Ref.~\citen{KT:2001,NS2},
whenever anomalous dimensions of all the chiral fields 
are uniquely determined at the fixed point,
one can use IR sum rules of the type \eq{sumrule} 
to show that each soft scalar mass is suppressed at IR.
In our case,
the symmetry of the theory ensures that
all the chiral fields have definite anomalous dimensions
at the new fixed point.
Actually the fixed point conditions of the couplings
$g_{{\rm SC}}$, $\lambda_\psi$ and $\lambda_S$ imply
\begin{eqnarray}
\frac{1}{2}\left(\gamma_\psi\right)_*
 =   \frac{1}{2}\left(\gamma_S\right)_*
 ={}-\frac{1}{2}\left(\gamma_{\Phi}+\gamma_{\bar{\Phi}}\right)_*
 =  \frac{3\Nc-\Nf}{\Nf}
 \equiv \gamma_* \ .
\label{gamma}
\end{eqnarray}
In this paper, we confine ourselves to $2\Nc<\Nf<3\Nc$
(\ie, the ``weakly-coupled side'' of the conformal window)
and treat $\gamma_*$ as a free parameter satisfying
\begin{eqnarray}
0 < \gamma_* < \frac{1}{2} \ .
\label{gamma:range}
\end{eqnarray}
We note that the higher dimensional operator 
$\left(\Phi\wb{\Phi}\right)^2$ remains irrelevant
as long as \Eq{gamma:range} is satisfied.

In our model, 
a mass term of SC-sector matter fields $\Phi$ and $\wb{\Phi}$
is generated through the first term in \Eq{W:dec}
once the singlet field develops a nonzero VEV $\VEV{S}$.
The mass parameter renormalized at a scale $\mu$
is given by $M_{\Phi}\fun{\mu}=\lambda_S\fun{\mu}\VEV{S\fun{\mu}}$,
in which $\lambda_S\fun{\mu}$ may be replaced
by its fixed point value $\lambda_{S*}$ of $\order{4\pi}$.
To determine the size of the VEV, 
we will analyze in \S\ref{sec:mechanism} the scalar potential
\begin{eqnarray}
 V\fun{S,\wb{S}}
  =  m^2_S\abs{S}^2
   - \left\{\frac{1}{3!}\,\kappa{}A_{\kappa}S^3
   + 
     {\rm H.c.}
     \right\}
   + \frac{1}{4}\abs{\kappa}^2\abs{S}^4
     \ ,
\label{V}
\end{eqnarray}
which includes soft SUSY-breaking parameters, $m^2_S$ and $A_\kappa$.

Some remarks are in order here.
The first remark concerns the physical definition of the decoupling scale.
It turns out that 
the VEV and thus the mass parameter $M_{\Phi}\fun{\mu}$ 
depend heavily on the renormalization scale $\mu$.
This is to be expected from the fact that
the fields involved here have large anomalous dimensions.
In terms of this running mass parameter, therefore, 
we define the decoupling scale $\Mc$ of SC sector 
in a self-consistent manner by
\begin{eqnarray}
\Mc = M_{\Phi}\fun{\mu=\Mc} \ .
\label{Mc:def}
\end{eqnarray}
The second remark is about SUSY breaking; 
We allow a generic set of soft SUSY-breaking parameters 
at the Planck scale $\MPl\approx{}10^{19}\,\GeV$.
Specifically we assume that 
the soft parameters are of a few times ${10}\,\TeV$.
This initial value appears rather large for $\GSM$-sector particles,
but actually,
their soft parameters at lower energy are reduced 
by a couple of SCFT effects:
The running gaugino mass of $\GSM$ is reduced toward IR 
through corrections by many extra $G_{\rm SM}$-chraged fields
before $\GSC$-sector decouples.
Moreover, SCFT dynamics forces the soft scalar masses
converge on a small value as we will explain in the next section.

\section{Renormalization Group Evolution}
\label{sec:RGE}

In this section,
we study renormalization group (RG) evolution of the parameters
that appear in the scalar potential \eq{V}.

The most important point in our scenario is 
power-law suppression of the cubic self-coupling $\kappa$ 
in the second term in the superpotential \eq{W:dec}.
As the first term in \Eq{W:dec} approaches the fixed point
$\lambda_S\rightarrow\lambda_{S*}$,
the singlet field $S$ obtains a large anomalous dimension
$\left(\gamma_S\right)_*=2\gamma_*$.
This fact has a significant effect on $\kappa$.
If we denote by $\Lambda$ the scale below which
the theory can be regarded as conformal,
we have a power-law suppression according to
\begin{eqnarray}
 \kappa\fun{\mu}
  \approx
 \kappa_0\left(\frac{\mu}{\Lambda}\right)^{3\gamma_*} \ ,
\label{eqn:kappa}
\end{eqnarray}
where $\kappa_0=\kappa\fun{\Lambda}=\order{1}$.
Later we will take $\Lambda=\MPl$ for simplicity.
[Namely, we will assume that
the $\lambda_\psi$ and $\lambda_S$
as well as $\GSC$ gauge coupling
are very close to the conformal fixed point
already at the Planck scale,
whereas we allow $\kappa_0=\kappa\fun{\Lambda}=\order{1}$.]

Another important point is IR suppression of 
soft SUSY-breaking terms \cite{SQCD_convergence,KT:2001}.
{}For instance,
the gaugino mass of $\GSC$ gauge theory
and $A$ parameters corresponding to the superpotential couplings
$\lambda_\psi$ and $\lambda_S$  become suppressed 
toward the superconformal fixed point.
Therefore we will neglect all of them hereafter.
Similarly the soft scalar masses $m^2_i$ ($i=\Phi,\wb{\Phi},\psi,S$)
are generally suppressed thanks to SCFT dynamics.
In fact, as we shall explicitly discuss shortly,
the IR suppression is violated by a small amount
due to nonvanishing gauge couplings $\alpha_a=g^2_a/(8\pi^2)$ 
and gaugino masses $m_a$ of $\GSM$ gauge theory:
\begin{eqnarray}
 m^2_i\fun{\mu}
 \longrightarrow
 m^2_{i(*)}\fun{\mu}
 = \order{\alpha_a{}m_a^2} \ ,
\label{eqn:mass}
\end{eqnarray}
where the index $a=\L,\R$ is for each factor group 
of $\GSM=SU(n)_\L\times{}SU(n)_\R$.
Moreover,
the `convergence value' $m^2_{i(*)}$ has only mild $\mu$-dependence
since $\GSM$-sector is weakly coupled
so that $\alpha_a$ and $m_a$ run very slowly.
In the following sections,
we will treat $m^2_{i(*)}$ as if it is a constant.

On the other hand, the $A_\kappa$ parameter corresponding 
to the cubic self-coupling $\kappa$ is not suppressed so much.
This is because the trilinear scalar coupling $\kappa A_\kappa$
receives power-law suppression 
in almost the same way as in \Eq{eqn:kappa}.
Accordingly, unlike the soft scalar mass parameter,
this $A$-parameter is approximately unchanged,
\begin{eqnarray}
 A_\kappa\fun{\mu}
 \approx
 A_\kappa\fun{\Lambda}e^{-\frac{\kappa_0^2}{6\gamma_*}}
 \equiv  A_{\kappa{}*} \ .
\label{eqn:A}
\end{eqnarray}
Thus we expect rather large $A$ term in the scalar potential \eq{V}.

Now, let us demonstrate IR convergence of soft scalar masses
via SCFT dynamics.
In our model, left-right symmetry implies
$m^2_{\Phi_\L}=m^2_{\Phi_\R}$ and
$m^2_{\bar\Phi_\L}=m^2_{\bar\Phi_\R}$
as well as $m^2_{\psi_\L}=m^2_{\psi_\R}$.
Therefore,
near the conformal fixed point,
RG equations for soft scalar masses take the form
\begin{eqnarray}
\mu\frac{d}{d\mu}\!
\begin{pmatrix}
m^2_\psi \cr m^2_\Phi+m^2_{\bar{\Phi}} \cr m^2_S
\end{pmatrix}
  =  {\cal M}
     \begin{pmatrix}
     m^2_\psi \cr m^2_\Phi+m^2_{\bar{\Phi}} \cr m^2_S
     \end{pmatrix}
    -\sum_{a=\L,\R}
     \begin{pmatrix}
     C_\psi^a \cr C_\Phi^a+C_{\bar{\Phi}}^a \cr C_S^a
     \end{pmatrix}
     4\alpha_a{}m^2_a \ .
\label{RGE:matrix}
\end{eqnarray}
The coefficient matrix ${\cal M}$ summarizes strongly-coupled effects 
on these soft scalar masses and is given by
\begin{eqnarray}
{\cal M}
  =  \begin{pmatrix}
      a_\psi & a_\psi & 0 \cr
      b_\psi & b_\psi+b_S+\Delta & b_S \cr
      0      & a_S               & a_S
     \end{pmatrix}
     \ , \quad
\Delta
  \equiv  \frac{2C_{{\rm SC}\,}\Nf\alpha_{{\rm SC}}^2}
               {1-\Nc\alpha_{{\rm SC}}} \ ,
\label{matrix}
\end{eqnarray}
where $a_i$ and $b_i$ ($i=\psi,S$) 
are coefficient functions of $\order{1}$.
At one-loop level, as we will explicitly work out in Appendix,
we find
\begin{eqnarray}
a_\psi = \frac{\Nc\abs{\lambda_\psi}^2}{8\pi^2} \ , \quad
a_S    = \frac{\Nf\Nc\abs{\lambda_S}^2}{8\pi^2} \ , \quad
b_\psi = \frac{\Nf\abs{\lambda_\psi}^2}{8\pi^2} \ , \quad
b_S    = \frac{2\abs{\lambda_S}^2}{8\pi^2} \ .
\label{coeff:oneloop}
\end{eqnarray}
All of these are of $\order{1}$ near the fixed point,
except that the last one $b_S$ is rather small.
We note also that the coefficients $a_i$ at one-loop level are equal to 
the anomalous dimensions $\left(\gamma_i\right)_*$ at the fixed point.

On the other hand,
the second term in RG equations \eq{RGE:matrix}
is perturbative corrections 
at the lowest order in gauge couplings $\alpha_a=g^2_a/(8\pi^2)$.
At one-loop level,
the coefficients $C$'s are just Casimir factors \eq{Casimir}
and thus $C_S^a=0$;
In general they are functions of SC-sector couplings
such as $g_{{\rm SC}}$ and $\lambda_i$ ($i=\psi,S$).

The coefficient matrix ${\cal M}$
is non-singular thanks to the $\Delta$-term,
$\det{\cal M}=a_\psi{}a_S\Delta$,
and we can write its inverse matrix in the form
\begin{eqnarray}
{\cal M}^{-1}
 =  \begin{pmatrix}
    a_\psi^{-1} & & \cr &0& \cr & & a_S^{-1}
    \end{pmatrix}
   +\frac{1}{\Delta}
    \begin{pmatrix}
    -1 \cr +1 \cr -1
    \end{pmatrix}
    \begin{pmatrix}
    \disp{-\frac{b_\psi}{a_\psi}} & +1 & \disp{-\frac{b_S}{a_S}}
    \end{pmatrix}
    \ .
\label{inverse}
\end{eqnarray}
Consequently we find that soft scalar masses converge in IR regime as
\begin{eqnarray}
\begin{pmatrix}
m^2_\psi \cr m^2_\Phi+m^2_{\bar{\Phi}} \cr m^2_S
\end{pmatrix}
&\longrightarrow& \sum_{a=\L,\R}
     \left[
     \begin{pmatrix}
      a_\psi^{-1}C_\psi^a \cr 0 \cr a_S^{-1}C_S^a
     \end{pmatrix}
    +\frac{r^a}{\Delta}
     \begin{pmatrix}
     -1 \cr +1 \cr -1
     \end{pmatrix}
     \right]_*4\alpha_a{}m^2_a \ ,
\label{mass:converge}
\end{eqnarray}
where we have defined
\begin{eqnarray}
r^a
 \equiv
     C_\Phi^a+C_{\bar{\Phi}}^a
    -\sum_{i=\psi,S}
\frac{b_i}{a_i}
\,C_i^a
 \ .
\label{ratio:def}
\end{eqnarray}
Note that
the scalar component of the $S$ could be tachyonic if $r^a>0$.
However, one-loop result \eq{coeff:oneloop} indicates that
this is unlikely (as long as $\Nf/\Nc>2$).
On the other hand, when $r^a<0$,
SC-matter fields $\Phi$ and $\wb{\Phi}$
have negative soft mass-squared,
but they will also have huge supersymmetric mass term \eq{Mc:def}.

In passing, we note that
the IR sum rules \eq{sumrule} corresponding to 
the superpotential couplings $\lambda_{\psi,S}$ 
get modified by non-conformal gauge interactions into
\begin{eqnarray}
m^2_\psi+m^2_\Phi+m^2_{\bar{\Phi}}
&\longrightarrow&
	\frac{1}{a_\psi}
	\sum_{a}4C_\psi^a\alpha_a{}m^2_a \ ,
\\
m^2_S+m^2_\Phi+m^2_{\bar{\Phi}}
&\longrightarrow&
	\frac{1}{a_S}
	\sum_{a}4C_S^a\alpha_a{}m^2_a \ .
\end{eqnarray}
At the lowest order in perturbative gauge coupling $\alpha_{a=\L,\R}$,
the correction to each sum rule is proportional to
the Casimir coefficient of SC-singlet field ($\psi$ or $S$).
On the other hand,
the correction to the sum rule corresponding to the $\Delta$-term
is proportional to the complicated combination $r$ defined above.

We see that
the convergence value of $m^2_S$ is more suppressed than,
or at most comparable to, squark/slepton mass $m^2_{\psi}$,
which is much smaller than $A^2_\kappa$.
Therefore we have a hierarchical relation
\begin{eqnarray}
 m^2_S\fun{\mu}
 \lsim  m^2_\psi\fun{\mu}
 \ll    A^2_\kappa\fun{\mu} \ ,
\label{approx}
\end{eqnarray}
for $\mu\ll\Lambda$.
Consequently, the trilinear coupling is more important 
in the scalar potential \eq{V} than soft scalar mass term.

\section{Generation of large mass scale}
\label{sec:mechanism}

Now we discuss generation of the decoupling scale \eq{Mc:def}
by analyzing the scalar potential \eq{V}.
In this section, we examine the tree potential
with the RG running parameters substituted.

Let us parameterize the scalar VEV as
$\VEV{S}=s e^{i\Theta}$ ($s\ge 0$).
The phase $\Theta$ is fixed by the $A$-term
which violates $U(1)_R$ symmetry.
In the following we take the couplings $\kappa$ and $A_\kappa$ 
to be real and positive, for simplicity.
As a result, the VEV is along $\Theta=0$ direction.
With this understanding, we write the scalar potential as
\begin{eqnarray}
 V\left(S=s e^{i\Theta}\right)
 &=&
    m^2_S\fun{\mu}s^2
  - \frac{1}{3}\,\kappa\fun{\mu}A_\kappa\fun{\mu}s^3
    \cos\left(3\Theta\right)
  + \frac{1}{4}\,\kappa^2\fun{\mu}\,s^4
    \ .
\label{V:running}
\end{eqnarray}
This potential has a nontrivial minimum
since the mass parameter is much smaller than the $A$-parameter.
In fact, the stationary condition
\begin{eqnarray}
 0
 \:\:=\:\:
 \left.\frac{\partial V\fun{s e^{i\Theta}}}{\partial s}
 \right|_{\Theta=0}
 &=&
 A^2_\kappa\fun{\mu}s\left[
  \frac{2m^2_S\fun{\mu}}{A^2_\kappa\fun{\mu}}-X+X^2
  \right]\ 
\end{eqnarray}
has solutions
\begin{eqnarray}
 X\equiv\frac{\kappa\left(\mu\right)}{A_\kappa\left(\mu\right)}\,s
 &=&
 \frac{1}{2}\left[
  1\pm\sqrt{1-\frac{8m_{S}^{2}\left(\mu\right)}
                   {A^2_\kappa\left(\mu\right)}}
 \right]
 \equiv\ x_{\pm} \ .
\end{eqnarray}
Note that for $\mu$ sufficiently smaller than $\Lambda$,
the values $x_{\pm}$ become almost independent of $\mu$.
In addition, if we use an approximation \eq{approx},
this expression simplifies to
\begin{eqnarray}
 x_{+}\ \approx\ 1 \ , \qquad
 x_{-}\ \approx\ \frac{2m^2_{S*}}{A^2_{\kappa{}*}} \ .
\end{eqnarray}
The potential takes the minimum at the point $x_{+}$
(while $x_{-}$ corresponds to a local maximum).
Therefore we find
\begin{eqnarray}
 \VEV{S}
 \:\:=\:\:
 x_{+}\,\frac{A_\kappa\fun{\mu}}{\kappa\fun{\mu}}
 \:\:\approx\:\:
 \frac{A_{\kappa{}*}}{\kappa_0}
 \left(\frac{\Lambda}{\mu}\right)^{3\gamma_*} \ ,
\label{VEV:tree}
\end{eqnarray}
where we have substituted
running parameters \eq{eqn:kappa} and \eq{eqn:A}.

An interesting point in \Eq{VEV:tree} is that
the scalar VEV is much enhanced from $A_{\kappa{}*}=\order{10}\,\TeV$.
Namely, as the superconformal regime persists longer and longer,
the self-coupling $\kappa\fun{\mu}$ becomes smaller and smaller.
As a result, the scalar potential becomes more flat
and the resultant VEV becomes larger.
Of course, power-law running as above should be stopped 
at the scale $\mu\approx\Mc$
where the SC-sector becomes massive and decouples.
Therefore, the above expression \eq{VEV:tree} is reliable 
only for a proper choice of $\mu\sim\VEV{S}$.
Actually a self-consistent definition \eq{Mc:def} 
of the decoupling scale leads to
\begin{eqnarray}
\Mc^{1+3\gamma_*}
\ =\ \left(\frac{x_{+}\lambda_{S*}}{\kappa_0}\right)
     A_{\kappa{}*}\Lambda^{3\gamma_*} \ .
\label{Mc:tree}
\end{eqnarray}
We see that the decoupling scale is enhanced
by powers of the UV scale $\Lambda$ below which
we have power-law running like \Eq{eqn:kappa}.

The generation of a large mass scale can be understood 
also in terms of RG-improved 
effective potential \cite{ColemanWeinberg,Kastening,improving}.
We will defer the discussion to the next section.

Let us illustrate our results numerically.
We take the parameters to be
$\kappa_0=1$, $A_{\kappa{}*}=10\,\TeV$ and $\Lambda=\MPl=10^{19}\GeV$.
We also assume $\lambda_{S*}=1$ and $m_{S*}/A_{\kappa{}*}=0.25$ 
for definiteness.
Then Figure~\ref{fig:dec3} shows
the size of the generated scale $\Mc$
as a function of $\gamma_*$ defined in \Eq{gamma}.
We see that the singlet VEV and thus the $\Mc$
can be as large as an intermediate mass scale
$\order{10^{12}{\rm -}10^{13}}\,\GeV$.
The decoupling scale is a bit larger for $\lambda_{S*}=\order{4\pi}$
as is expected from naive dimensional analysis.
Note also that
we would have even larger $\Mc$ 
if $\kappa_0$ is suppressed for some reasons.

\begin{figure}[tb]
\begin{minipage}{0.475\textwidth} 
\begin{center}
  \includegraphics[width=0.95\textwidth]{fig_dec3.ai}
  \caption{
	The decoupling scale $M_{\rm c}$
	as a function of anomalous dimension $\gamma_{\ast}$
	in the case of cubic superpotential ($p=3$).
  } 
  \label{fig:dec3}
\end{center}
\end{minipage}
\hspace*{0.03\textwidth}
\begin{minipage}{0.475\textwidth} 
\begin{center}
  \includegraphics[width=0.95\textwidth]{fig_dec4.ai}
  \caption{
	The decoupling scale $M_{\rm c}$
	as a function of anomalous dimension $\gamma_{\ast}$
	in the case of quartic superpotential ($p=4$).
  } 
  \label{fig:dec4}
\end{center}
\end{minipage}
\end{figure}

Up to here, we have considered the superpotential 
with a cubic self-interaction, \Eq{W:dec}.
Let us make a brief comment on the case with higher order coupling:
\begin{eqnarray}
W_{{\rm dec}}^{(p)}
 &=& \lambda_{S\,}S\leftB[
     \Phi_\L\wb{\Phi}_\R
    +\Phi_\R\wb{\Phi}_\L
     \rightB]
    +\frac{1}{p!}\,\frac{\kappa_p}{\MPl^{p-3}}\,S^p \ .
\end{eqnarray}
If we consider larger $p$ ($>3$), we can obtain larger VEV
and thus larger decoupling scale of SCFT.
{}For instance,
Figure~\ref{fig:dec4} shows the result of $p=4$ case.
The parameters are the same as before.
In this case, the VEV for $\gamma_*=0$ is already 
of order of $10^{12}\,\GeV$,
which can be enhanced up to $10^{15}\,\GeV$
by the effect of large anomalous dimension.
In general,
a phenomenologically favored value \cite{NS1,KT:2001,NS2},
$\Mc=\order{10^{13}{\rm -}10^{15}}\,\GeV$,
can easily be obtained.

\section{RG-improved potential and enhanced mass scale}
\label{sec:improving}

Although there is nothing wrong in the above discussion
using the tree potential,
it is a bit puzzling that
the minimum \eq{VEV:tree} of the tree potential
depends on the renormalization scale $\mu$
in a way which differs from the scaling
\begin{eqnarray}
S\fun{\mu{}e^t} = S\fun{\mu}e^{-\gamma_*t} \ .
\label{scaling:mass}
\end{eqnarray}
To clarify this point,
let us discuss RG-improvement of the tree potential.

Generally RG improvement of the effective potential $V\fun{S}$
means \cite{ColemanWeinberg,Kastening,improving} that
one sums up corrections to $V\fun{S}$ systematically 
by requiring that it satisfies the RG equation
\begin{eqnarray}
0&=& \mu\frac{d}{d\mu}
     V\leftB[S\fun{\mu},m^2_S\fun{\mu},\kappa\fun{\mu},
	     A_\kappa\fun{\mu},\lambda_S\fun{\mu};\,\mu
     \rightB]
\nonumber\\
 &=& \left[
     \mu\frac{\partial}{\partial\mu}
    +\beta_{m^2_S}\frac{\partial}{\partial{m^2_S}}
    +\left\{
     \beta_\kappa{}\frac{\partial}{\partial\kappa}
    +\beta_{A_\kappa}\frac{\partial}{\partial{A_\kappa}}
    +\beta_{\lambda_S}\frac{\partial}{\partial{\lambda_S}}
    -
     \frac{\gamma_S}{2}
     S\frac{\partial}{\partial{S}}
    +
     {\rm H.c.}
     \right\}
     \right]\!{}V  
     \,. \qquad
\label{V:RGE}
\end{eqnarray}
Since the complex phase is irrelevant for our present purpose,
we will treat the couplings $\lambda_S$, $\kappa$, $A_\kappa$
as well as $S$ as if they are real quantities.
[Then we just drop Hermitian conjugated terms in \Eq{V:RGE}.]
{}Furthermore,
we will neglect the soft mass term for a moment,
since we are mainly interested in large $S$ region.
Then the general solution reads
\begin{eqnarray}
 & &
 V\leftB[S\fun{\mu},\kappa\fun{\mu},
	 A_\kappa\fun{\mu},\lambda_S\fun{\mu};\,\mu
  \rightB]
\nonumber\\
 &=& 
 V\leftB[S\fun{\mu{}e^t},\kappa\fun{\mu{}e^t},
	 A_\kappa\fun{\mu{}e^t},\lambda_S\fun{\mu{}e^t};\,\mu{}e^t
  \rightB] \ .
\label{V:RGE:sol}
\end{eqnarray}
In the superconformal regime we are considering,
only nontrivial RG evolutions are given by \Eq{scaling:mass} and
\begin{eqnarray}
\kappa\fun{\mu{}e^t} = \kappa\fun{\mu}e^{3\gamma_*t} \ .
\label{RGE:SC}
\end{eqnarray}
One can reliably calculate RHS of \Eq{V:RGE:sol} 
by making a clever choice of $t$.
{}Following Ref.~\citen{improving},
we choose $t$ according to
\begin{eqnarray}
\mu^2{}e^{2t} = M^2_{\Phi}\fun{\mu{}e^t} \ ,
\label{improving:general}
\end{eqnarray}
where $M_{\Phi}\fun{\mu}$ is 
the mass parameter of $\Phi$ and $\wb{\Phi}$,
renormalized at a scale $\mu$,
in the scalar background $S=\VEV{S}$.
{}For large $S$ region,
we can neglect SUSY-breaking contribution
to $M^2_{\Phi}\fun{\mu}$, so that we have
$\mu{}e^{t}=\lambda_S\fun{\mu}S\fun{\mu}e^{-\gamma_*t}$, \ie,
\begin{eqnarray}
e^{\gamma_*{t}}
  =  \left[\frac{\lambda_S\fun{\mu}S\fun{\mu}}{\mu}
     \right]^{\frac{\gamma_*}{1+\gamma_*}}
  \equiv \sqrt{Z\!\left[S\fun{\mu};\,\mu\right]}
     \ .
\label{improving:quartic}
\end{eqnarray}
{}For $S$ smaller than $A_\kappa$,
we should properly take care of the existence of 
multi-mass scales \cite{improving},
but we will not attempt it here.

Now we apply the above prescription
for improving the tree potential
\begin{eqnarray}
 V\leftB[S\fun{\mu},\kappa\fun{\mu},A_\kappa\fun{\mu};\,\mu\rightB]
 &=&{}-\frac{1}{3}\,
       \kappa\fun{\mu{}e^t}A_\kappa\fun{\mu{}e^t}S^3\fun{\mu{}e^t}
      +\frac{1}{4}\,
       \kappa^2\fun{\mu{}e^t}S^4\fun{\mu{}e^t}
\nonumber\\
 &=&{}-\frac{1}{3}\,
       \kappa\fun{\mu}A_\kappa\fun{\mu}S^3\fun{\mu}
      +\frac{1}{4}\,
       \kappa^2\fun{\mu}S^4\fun{\mu}e^{2\gamma_*t}
       \ .
\label{V:tree}
\end{eqnarray}
Evaluating the last expression by using \Eq{improving:quartic} gives
\begin{eqnarray}
 V\leftB[S\fun{\mu};\,\mu\rightB]
\ ={}-\frac{1}{3}\,
      \kappa\fun{\mu}A_\kappa\fun{\mu}S^3\fun{\mu}
     +\frac{1}{4}\,
      \kappa^2\fun{\mu}S^4\fun{\mu}
      \left[\frac{\lambda_S\fun{\mu}S\fun{\mu}}{\mu}
      \right]^{\frac{2\gamma_*}{1+\gamma_*}}
      \ . \qquad
\label{V:tree:improved}
\end{eqnarray}
By construction,
this potential satisfies the RG equation.

We can also incorporate the soft scalar mass term 
as follows.
As we mentioned below \Eq{eqn:mass},
the soft mass parameter $m^2_S\fun{\mu}$ for $\mu\ll\Lambda$ 
converges on a small value $m^2_{S*}$
which may be treated as a constant.
Then, by repeating the above procedure, we find
the RG-improved tree potential including the mass term to be
\begin{eqnarray}
V\leftB[S\fun{\mu};\,\mu\rightB]
 &=& \frac{m^2_{S*}S^2\fun{\mu}}{Z\!\left[S\fun{\mu};\mu\right]}
    -\frac{1}{3}\,
     \kappa\fun{\mu}A_\kappa\fun{\mu}S^3\fun{\mu}
    +\frac{1}{4}\,
     \kappa^2\fun{\mu}S^4\fun{\mu}
     Z\!\left[S\fun{\mu};\mu\right]
     \ , \qquad
\label{V:RGE:final}
\end{eqnarray}
where the function $Z\left[S\fun{\mu}\right]$ is defined
in \Eq{improving:quartic}.

The minimum of the potential \eq{V:RGE:final}
can be found in quite the same way as before.
The stationary condition $S\left(\partial{V}/\partial{S}\right)=0$
leads to
\begin{eqnarray}
     \frac{\kappa\fun{\mu}S\fun{\mu}Z\!\left[S\fun{\mu};\,\mu\right]}
          {A_\kappa\fun{\mu}}
 &=& \frac{1+\gamma_*}{2+3\gamma_*}
     \left[
     1\pm \sqrt{
     1-\frac{2+3\gamma_*}{\left(1+\gamma_*\right)^2}
     \left(\frac{4m^2_{S*}}{A^2_{\kappa{}*}}\right)
     }
     \right]
  \equiv  x'_{\pm}
     \ . \quad
\label{solution:improved}
\end{eqnarray}
Thus the minimum is determined by
\begin{eqnarray}
     \left[\frac{\lambda_S\fun{\mu}S\fun{\mu}}{\mu}
     \right]^{\frac{1+3\gamma_*}{1+\gamma_*}}
  =  \frac{x'_{+}\lambda_S\fun{\mu}A_\kappa\fun{\mu}}
          {\kappa\fun{\mu}\mu} \ ,
\nonumber
\end{eqnarray}
or equivalently,
\begin{eqnarray}
M_{\Phi}\fun{\mu}
 \equiv
     \lambda_{S}\fun{\mu}S\fun{\mu}
  =  \left(\frac{x'_{+}\lambda_{S*}A_{\kappa{}*}}{\kappa\fun{\mu}}
     \right)^{\frac{1+\gamma_*}{1+3\gamma_*}}
     \mu^{\frac{2\gamma_*}{1+3\gamma_*}}
     \ .
\label{VEV:tree:improved}
\end{eqnarray}

\begin{figure}[tb]
\begin{center}
  \includegraphics[width=0.5\textwidth]{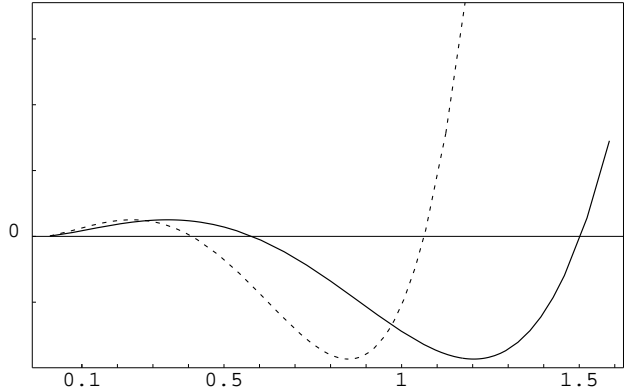}
  \caption{
	RG-improved potentials with 
	$\mu=1.0\times{}10^{13}\,\GeV$ (dashed line) and
	$\mu=0.5\times{}10^{13}\,\GeV$ (solid line).
	We take $\gamma_*=1/2$ and
	$m_{S*}/A_{\kappa{}*}=0.25$ for definiteness.
	The horizontal axis is $S\fun{\mu}$ 
	in the unit of $10^{13}\,\GeV$.
  } 
  \label{fig:mudep}
\end{center}
\end{figure}

Although the RG-improved potential satisfies the RG equation, 
its minimum \textit{does} depend on the renormalization scale $\mu$,
as is seen from Fig.~\ref{fig:mudep}.  In fact, we have
\begin{eqnarray}
M_{\Phi}\fun{\mu}
  =  \left[
     \left(C'A_{\kappa{}*}\right)^{1+\gamma_*}
     \Lambda^{2\gamma_*}
     \right]^{\frac{1}{1+3\gamma_*}}
     \left(\frac{\mu}{\Lambda}\right)^{{}-\gamma_*}
     \ .
\label{runningmass:improved}
\end{eqnarray}
where $C'\equiv{}x'_{+}\lambda_{S*}/\kappa_0$ is a numerical coefficient.
This is precisely the scaling of the running mass parameter
of $\Phi$ and $\wb{\Phi}$.
In this way, we find that
the scaling \eq{scaling:mass} is correctly reproduced
by our RG-improved tree potential,
contrary to the naive tree potential \eq{VEV:tree}.

The physical decoupling scale,
defined through the relation \Eq{Mc:def},
is essentially unchanged from the previous result \eq{Mc:tree}.
This fact may be seen most easily as follows:
If we use in \Eq{solution:improved} 
the self-consistent definition \eq{Mc:def},
the `wavefunction' factor $Z\!\left[S\fun{\mu};\,\mu\right]$
becomes trivial.
Actually from the above expression \eq{VEV:tree:improved}
for the running mass parameter, we obtain 
\begin{eqnarray}
\Mc
\ =\ \left[C'A_{\kappa{}*}\Lambda^{3\gamma_*}
     \right]^{\frac{1}{1+3\gamma_*}}
     \ ,
\label{Mc:final}
\end{eqnarray}
which agrees with \Eq{Mc:tree} up to order one factor.

\begin{figure}[tb]
\begin{center}
  \includegraphics[width=0.5\textwidth]{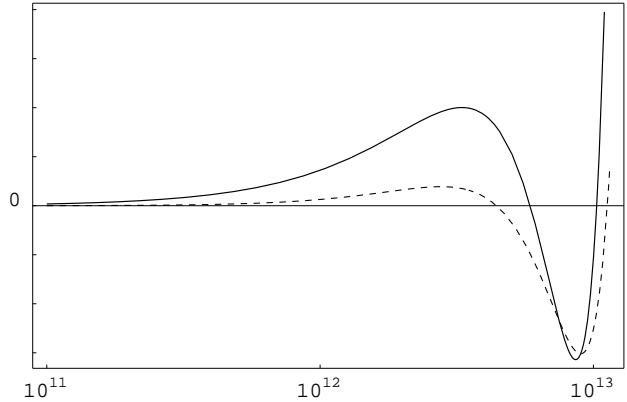}
  \caption{
	The RG-improved tree potential (solid line) 
	versus the tree potential 
	with running parameters substituted (dashed line).
	Both are renormalized at $\mu=\Mc$ given by \Eq{Mc:final}.
	The horizontal axis stands for 
	the value of $S$ [GeV] in log scale.
  }
  \label{fig:compare}
\end{center}
\end{figure}

{}Figure~\ref{fig:compare} shows a comparison between
the potentials with and without RG improvement,
evaluated at the physical mass scale \eq{Mc:final}.
The parameters are the same as before, 
except that we take a larger value of $m_{S*}/A_{\kappa{}*}=0.3$
to highlight the behaviour in small $S$ region.
We see that the minimum almost coincides with each other.
On the other hand,
the shape of the potential is slightly changed
as a result of the RG improvement.
The mass term is enhanced for small $S$ region
while the quartic term is enhanced for large $S$ region
due to $S$-dependent $Z$ factor.

{}Finally, 
we should note that the effective potential \eq{V:RGE:final}
receives nontrivial improvements
only through a field-dependent `wavefunction' factor 
$Z\!\left[S\fun{\mu};\,\mu\right]$ of the scalar field $S\fun{\mu}$.
Of course, this property is consistent 
with non-renormalization theorem of the superpotential.
In the present case,
that property is true \textit{even in the presence of SUSY breaking}\/.
This is not only because
we have neglected SUSY-breaking contributions
in \Eq{improving:general},
but also because
soft SUSY-breaking parameters become almost constant 
after RG evolution in superconformal regime.
Consequently
the potential can be trusted only for energy scale
larger than SUSY-breaking parameters
and sufficiently smaller than $\Lambda$.

\section{Conclusion and Discussion}
\label{sec:conclusion}

We have shown that
a large mass scale can be generated
if we consider a proper coupling of a gauge-singlet field to 
a strongly-coupled gauge dynamics with superconformal fixed point.
Such SCFT dynamics was proposed before as 
a solution to SUSY flavor problem even in the presence of 
generic type of soft SUSY-breaking parameters.
The generated mass scale plays a role of decoupling scale of the SCFT
and can be as large as the intermediate scale.
An interesting point of our mechanism is that
the decoupling scale of the SCFT is enhanced from 
the scale of soft SUSY-breaking parameters
thanks to its own strongly-coupled dynamics.

The generation of large scalar VEV can be understood
by suppression of its self-coupling 
if we just substitute the running couplings into the tree potential;
Alternatively it can be understood,
if we consider RG-improved effective potential,
by enhancement of the scalar VEV 
owing to large anomalous dimension of the scalar field.
In both cases, the decoupling scales are calculated
in a self-consistent manner,
and the results coincide with each other
(modulo a numerical factor of order one).

There remain some unsettled issues.

We have assumed that the singlet field $S$ 
has no mass term in the superpotential \eq{W:dec}.
The coupling of $S$ to SM particles should also be absent
or highly suppressed, not to generate unwanted huge mass terms.
These properties could be explained on symmetry ground,
but it would require detailed model building
which we will not attempt here.
Note that our mechanism will work even if the single field $S$
has a supersymmetric mass of $\TeV$ scale
as long as the trilinear term still dominates it.

The masses of the fluctuation modes of the $S$ field
are typically of $A_{\kappa{}*}=\order{10}\,\TeV$.
To determine their precise interactions to other particles
would require the precise knowledge of higher dimensional
interactions.
Here we only note that
such interaction, if exists, will also receive
a suppression due to large anomalous dimension.

The RG-improved potential in the last section can be well justified 
for $\gamma_*\ll{}1$.
Strictly speaking, however, the correct treatment of RG improvement
in strongly-coupled theories is not known
and is beyond the scope of the present paper.
We have also neglected perturbative corrections 
from weakly-coupled $\GSM$-sector.
With these corrections, soft SUSY-breaking parameters 
are no longer constant after IR convergence,
but we do not expect that 
our result will be modified substantially.

{}Finally we add a remark on model building. Although 
the powers in the formula \eq{Mc:final} for $\Mc$ is determined 
by the value $\gamma_*$ of the anomalous dimension 
at the IR stable fixed point,
the size of $\Mc$ is not completely UV insensitive;
it does depend on high-energy values of the self-coupling $\kappa$
and the corresponding $A$-parameter.
This is a unpleasant feature of our mechanism.
{}For instance, 
let us suppose that
if SC-sector gauge group consists of several factor groups,
$\GSC=\prod_{k}\GSC^{(k)}$ ,
and each factor group needs its own decoupling sector.
Then each $\GSC^{(k)}$ theory would decouple at different scale,
invalidating the suppression of SUSY flavor violation.
Therefore, SC sector with a simple gauge group is suitable
for our decoupling mechanism 
not to reintroduce non-degeneracy of sfermion masses.
A concrete model building would be discussed elsewhere.

\section*{Acknowledgements}

The authors thank the Yukawa Institute for Theoretical Physics at
Kyoto University, where a portion of this work was carried out during
the YITP-W-04-04 on ``Progress in Particle Physics" and 
the YITP-W-04-08 on ``Summer Institute 2004".
T.~K.\ is supported in part by the Grants-in-Aid for Scientific Research
(No.~16028211) 
and the Grant-in-Aid for the 21st Century COE
``The Center for Diversity and Universality in Physics"
from the Ministry of Education, Science, Sports and Culture, Japan.
H.~N.\ and H.~T.\ 
are supported in part by the Grants-in-Aid for Scientific Research
(No.~16540238 and No.~13640272, respectively)
from the Ministry of Education, Science, Sports and Culture, Japan.

\appendix
\section{IR Convergence of Soft Scalar Mass}
\label{sec:convergence}

In this appendix we study RG equations of soft parameters 
to some details.

We are particularly interested in RG evolution of soft scalar masses
$m^2_\Phi+m^2_{\bar{\Phi}}$, $m^2_\psi$ and $m^2_S$
near the superconformal fixed point.
To see this, let us recall general form of RG equations
for soft scalar masses \cite{softbeta,softbeta:delta,Terao};
Given anomalous dimension matrix $\gamma^i{}_j$ as a function of
gauge couplings $\alpha_a\equiv{}g^2_a/(8\pi^2)$ 
as well as superpotential couplings
$\left(y^{ijk}/3!\right)\phi_i\phi_j\phi_k$
and $y_{\ell{}mn}=(y^{\ell{}mn})^*$,
RG equations for the corresponding soft scalar mass $(m^2)^i{}_j$
can be calculated according to the formula
\begin{eqnarray}
\mu\frac{d}{d\mu}(m^2)^i{}_j
  =  D_2\gamma^i{}_j\fun{\alpha,y,\wb{y}} \ .
\end{eqnarray}
Here $D_2$ is a differential operator defined by
\def\l{\ell}
\begin{eqnarray}
D_2
 &=& \wb{D}_1 D_1 
    +X_a\alpha_a\frac{\partial}{\partial \alpha_a}
    +\frac{1}{3!}\,
     X^{ijk}_{\l{}mn}\,\frac{1}{2}
     \left(y^{\l{}mn}\frac{\partial}{\partial y^{ijk}} 
          +y_{ijk}\frac{\partial}{\partial y_{\l{}mn}}\right) \ ,
\label{D2}\\[3pt]
D_1
 &=& m_a\alpha_a\frac{\partial}{\partial\alpha_a}
    -\frac{1}{3!}\left(yA\right)^{ijk}
     \frac{\partial}{\partial y^{ijk}} \ , 
\end{eqnarray}
where $m_a$ is gaugino mass corresponding to $\alpha_a$, and
\begin{eqnarray}
X^{ijk}_{\l{}mn}
 &\equiv& (m^2)^{i}_{\ \l\,}\delta^{j}_{m\,}\delta^{k}_{n}
         +\delta^{i}_{\l\,}(m^2)^{j}_{\ m\,}\delta^{k}_{n}
         +\delta^{i}_{\l\,}\delta^{j}_{m\,}(m^2)^{k}_{\ n} \ ,
\\[5pt]
X_a\ 
 &\equiv& \abs{m_a}^2+\Delta_a \ , \quad
\Delta_a^{{\rm NSVZ}}
  ={}-\frac{\alpha_a}{1-T_a\alpha_a}
      \left[\sum_i{T_i}m^2_i-T_a\abs{m_a}^2\right] \ .      
\end{eqnarray}
In the last equation,
$T_i\equiv{}T\fun{R_i}$ is Dynkin index
and $T_a\equiv{}C_2\fun{G_a}$ is quadratic Casimir.
The $\Delta_a$ term, 
whose necessity was noticed in Ref.~\citen{softbeta:delta},
is written in a scheme of Ref.~\citen{NSVZ}.

To be concrete,
let us explicitly work out RG equations for our model 
by using anomalous dimensions at one-loop level;
\begin{eqnarray}
\gamma_{\Phi}+\gamma_{\bar{\Phi}}
 &=& \frac{\Nf\abs{\lambda_{\psi}}^2}{8\pi^2}
    +\frac{2\abs{\lambda_S}^2}{8\pi^2}
    -\sum_{a=\L,\R}2\left(C^a_{\Phi}+C^a_{\bar\Phi}\right)\alpha_a
    -4C_{{\rm SC}\,}\alpha_{{\rm SC}} \ , \qquad
\nonumber\\
\gamma_\psi
 &=& \frac{\Nc\abs{\lambda_\psi}^2}{8\pi^2}
    -\sum_{a=\L,\R}2C^a_{\psi}\alpha_a \ , \qquad
\gamma_S
  =  \frac{\Nf\Nc\abs{\lambda_S}^2}{8\pi^2} \ ,
\end{eqnarray}
where we have neglected self-coupling $\kappa$ of $S$.
Left-right symmetry which exchanges $SU(n)_\L$ and $SU(n)_\R$ 
implies $\alpha_\L=\alpha_\R$.
Nonvanishing Casimir factors are given by
\begin{eqnarray}
C_\psi^{a=\L, \R}
  =  C_{\Phi_\L}^{a=\L}
  =  C_{\Phi_\R}^{a=\R}
  =  \frac{n^2-1}{2n}
\label{Casimir}
\end{eqnarray}
for $\GSM=SU(n)_\L\times{}SU(n)_\R$, and 
$C_{{\rm SC}}=\left(\Nc^2-1\right)/\left(2\Nc\right)$ 
for $\GSC=SU(\Nc)$.

Since left-right symmetry implies
$m^2_{\Phi_\R}=m^2_{\Phi_\L}$ and
$m^2_{\bar\Phi_\R}=m^2_{\bar\Phi_\L}$, we have
\begin{eqnarray}
\mu\frac{dm^2_S}{d\mu}
 &=& \frac{\Nf\Nc\abs{\lambda_S}^2}{8\pi^2}
     \left(m^2_S+m^2_{\Phi}+m^2_{\bar{\Phi}}\right) \ ,
\\
\mu\frac{dm^2_\psi}{d\mu}
 &=& \frac{\Nc\abs{\lambda_\psi}^2}{8\pi^2}
     \left(m^2_\psi+m^2_{\Phi}+m^2_{\bar{\Phi}}\right)
    -\sum_{a=\L,\R}4C_\psi^a\alpha_a{}m_a^2 \ ,
\label{RGE:SCsum}\\
\mu\frac{d}{d\mu}\!\left(m^2_\Phi+m^2_{\bar{\Phi}}\right)
 &=& \frac{\Nf\abs{\lambda_\psi}^2}{8\pi^2}
     \left(m^2_\psi+m^2_{\Phi}+m^2_{\bar{\Phi}}\right)
    +\frac{2\abs{\lambda_S}^2}{8\pi^2}
     \left(m^2_S+m^2_{\Phi}+m^2_{\bar{\Phi}}\right)
\nonumber\\
 & & \!\!\!{}
    +\frac{2\Nf{}C_{{\rm SC}\,}\alpha_{{\rm SC}}^2}
          {1-\Nc\alpha_{{\rm SC}}}
     \left(m^2_{\Phi}+m^2_{\bar{\Phi}}\right)
    -\sum_{a=\L,\R}4\left(C_\Phi^a+C_{\bar{\Phi}}^a\right)
                   \alpha_a{}m_a^2 \ .\qquad
\end{eqnarray}
Note that soft scalar masses appear in RG equations 
through the combinations
$m^2_\psi+m^2_{\Phi}+m^2_{\bar{\Phi}}$
and 
$m^2_S+m^2_{\Phi}+m^2_{\bar{\Phi}}$,
except the $\Delta$-term which involves the soft masses, 
$m^2_{\Phi}+m^2_{\bar{\Phi}}$, of SC-sector matter fields only.
Therefore, when we write the above RG equations 
in the matrix form \eq{RGE:matrix},
the coefficient matrix \eq{matrix} becomes singular 
in the limit $\Delta\rightarrow0$.
This is one reason why we have kept the $\Delta$ term.
Another reason is that
although this term is apparently subleading in loop expansion,
it is a leading order term in large $\Nc\sim\Nf$ sense.

\end{document}